\documentclass[twocolumn,prb,showpacs]{revtex4}
\usepackage{graphics,epsf}

\newcommand{\xy}{\textsl{XY} }
\newcommand{\unitx}{\hat{\mathbf{x}} }
\newcommand{\unity}{\hat{\mathbf{y}} }
\newcommand{\irms}{I_\mathrm{rms} }

\begin{document}

\title{Phase Transitions in
  the Two-Dimensional Random Gauge \xy Model}
\author {Petter \surname{Holme}}
\email{holme@tp.umu.se}
\affiliation {Department of Physics, Ume{\aa}
University, 901 87 Ume{\aa}, Sweden}
\author {Beom Jun \surname{Kim}}
\email{beomjun@ajou.ac.kr}
\affiliation{Department of Molecular Science
  and Technology, Ajou University, Suwon 442-749, Korea}
\author {Petter \surname{Minnhagen}}
\email{minnhagen@nordita.dk}
\affiliation {NORDITA, Blegdamsvej 17, DK-2100, Copenhagen, Denmark}
\affiliation {Department of Physics, Ume{\aa}
University, 901 87 Ume{\aa}, Sweden}

\begin{abstract}
The two-dimensional random gauge \xy model, where the quenched
random variables are magnetic bond angles uniformly distributed
within $[-r\pi, r\pi]$ ($0 \leq r \leq 1$), is studied via Monte
Carlo simulations. We investigate the phase diagram in the plane
of the temperature $T$ and the disorder strength $r$, and infer,
in contrast to a prevailing conclusion in many earlier studies,
that the system is superconducting at any disorder strength $r$
for sufficiently low $T$. It is also argued that the
superconducting to normal transition has different nature at weak
disorder and strong disorder: termed Kosterlitz-Thouless (KT) type
and non-KT type, respectively. The results are compared to earlier
works.
\end{abstract}



\maketitle
\section{Introduction}
The \xy gauge glass model~\cite{huse} has attracted much interest
in connection to the vortex glass phase of high-$T_c$
superconductors~\cite{fisher}. In three dimensions (3D) there is a
general consensus that the \xy gauge glass model exhibits a
finite-temperature glass transition~\cite{huse,3d,noT0,simkin}.
However, in 2D there exist conflicting evidences: On the one hand,
equilibrium studies of defect
energy~\cite{3d,simkin,other:T0:studies,FTY}, Monte Carlo (MC)
simulations of the root-mean-square current~\cite{reger}, and
resistance calculation~\cite{hyman} have suggested that no
finite-$T$ ordering exists. The glass order parameter has
furthermore been analytically shown to vanish at any finite
$T$~\cite{nishimori}. On the other hand, the MC studies of the
glass susceptibility~\cite{choi}, as well as dynamical simulations
of the resistance~\cite{li,bjk} and non-equilibrium
relaxation~\cite{bjk1}, have indicated a possibility of a
finite-$T$ transition. Also the MC simulations of the helicity
modulus in Ref.~\cite{jeon} were interpreted as being compatible
with a finite-$T$ transition.

In this paper we study a generalization of the 2D \xy gauge glass
model---the random gauge \xy model---where both the temperature
$T$ and disorder strength $r$ can be varied. When $r$ has the
maximum value 1, it corresponds to the usual \xy gauge glass
model, while in the opposite limit of $r=0$, the standard \xy
model without disorder is recovered~\cite{jeon,maucourt,more}. It
has been proposed that there is a Kosterlitz-Thouless (KT) like
transition at a finite temperature $T_c$ when the disorder
strength is sufficiently small, and that as $r$ is increased $T_c$
becomes smaller until it vanishes as $r$ reaches the critical
disorder strength $r_c$ ($<1$) and $T_c = 0$ for $r >
r_c$~\cite{maucourt,more}. However, even if the glass order
parameter is zero and even if there is no finite-$T$ KT transition
for $r>r_c$, the existence of a finite-$T$ transition with a
different character cannot \textit{a priori} be ruled out.

\begin{figure}
\centering{{\resizebox*{!}{4.5cm}{\includegraphics{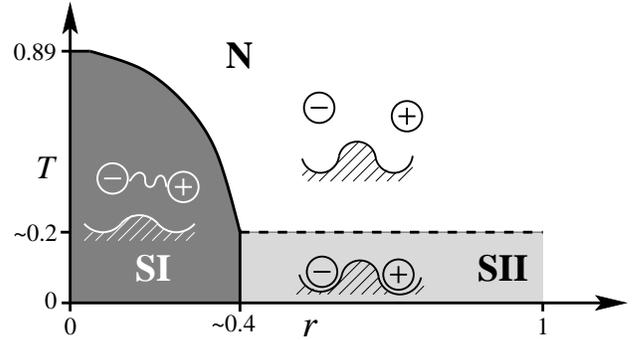}}}}
\caption{Sketch of the phase diagram in the disorder strength $r$ and the
temperature $T$ (in units of $J/k_B$ with the coupling strength $J$) plane.
There exist two superconducting phases at low temperatures: the low-$T$ KT
phase marked by ``SI'' and the distinct phase ``SII'' for small $r$ and large
$r$, respectively. The high temperature normal phase is marked as ``N''.  The
solid phase boundary represents the boundary of the low-$T$ KT phase whereas
the dashed represents the normal to superconducting transition for larger $r$.
}
\label{PhD}
\end{figure}

In the present paper we perform extensive MC simulations to study
the phase transition of the random gauge $XY$ model. It is found
that the system is superconducting at any $r$ and that the
transition from normal to superconducting phase is consistent with
a Kosterlitz-Thouless (KT) type at weak disorder and a non-KT type
at strong disorder. In addition to this we suggest that there
exist two different superconducting phases at sufficiently low
temperatures separated by a non-KT phase transition (see
Fig.~\ref{PhD}). For the special case of $r=1$, which corresponds
to the 2D \xy gauge glass model, we also compute the
root-mean-square current in the same way as in Ref.~\cite{reger}
and, in contrast to Ref.~\cite{reger}, again consistently find a
finite-temperature transition.

\section{Model and Simulations}
The Hamiltonian of the 2D random gauge \xy model on an
$L\times L$ square lattice under the fluctuating twist
boundary condition (FTBC)~\cite{olsson} is given by
\begin{equation} \label{eq_H}
 \hat{H}=- J \sum_{\langle ij\rangle}
\cos\left(\phi_{ij}\equiv\theta_i-\theta_j- \frac{1}{L}\mathbf{r}_{ij}\cdot
\mathbf{\Delta} - A_{ij}\right),
\end{equation}
where $J$ is the coupling strength (set to unity from now on), the
sum is over nearest neighbor pairs, $\mathbf{ r}_{ij} \equiv
\unitx$ ($\unity$) if $j = i+\unitx (\unity)$. The phase angle
$\theta_i$ at the lattice point $i$ satisfies the periodicity
$\theta_{i+L\unitx} = \theta_{i+L\unity} = \theta_i$, and
$A_{ij}\in r[-\pi,\pi]$ is a uniform quenched random variable with
the disorder strength $0\leq r\leq 1$. The twist variable
$\mathbf{ \Delta} = (\Delta_x, \Delta_y)$ corresponds to the
global twist across the system, i.e., the summation of the gauge
invariant phase difference $\phi_{ij}$ along the $x$ ($y$)
direction equals $\Delta_x$ ($\Delta_y$). For a given disorder
realization, we first compute the distribution $P( \mathbf{
\Delta} )$, which is related to the free energy $F$ by $\partial
F/\partial \mathbf{ \Delta} = -T(\partial \ln P/\partial \mathbf{
\Delta})$. The twist variable ${\bf\Delta}_0$ which minimizes $F$
(or maximizes $P$) is determined from $P$ and then fixed when the
helicity modulus $\Upsilon \equiv \partial^2 F/\partial \Delta^2$
and the 4th order modulus $\Upsilon_4 \equiv
\partial^4 F/\partial \Delta^4$ are computed.

To ensure that the cooling is slow enough, we simultaneously cool
two replicas ($\alpha$ and $\beta$) of the system and measure
${\bf\Delta}_0^{\alpha,\beta}$. For the first cooling at a new
temperature we use 120000 update sweeps (for spin and twist
variables respectively). Then we check that
\begin{equation}\label{eq:annealing}
|\Delta_{x,0}^\alpha-\Delta_{x,0}^\beta| <\delta \mbox{~and~}
|\Delta_{y,0}^\alpha-\Delta_{y,0}^\beta| <\delta
\end{equation}
where $\delta$ sets the precision of the cooling. The idea with the
annealing condition Eq.~(\ref{eq:annealing}) is to keep the system close
to the lowest-energy state at a particular temperature (so the system
does not freeze into a local minimum that bias $[\Upsilon]$ and
$[\Upsilon_4]$). Since the system moves more swiftly over the
configuration space the higher the temperature is we can choose
$\delta$ increasing with temperature. A choice that proves good in
practice is 
\begin{equation}\label{eq:delta}
  \delta=\left\{\begin{array}{ll} 0.02\pi & \mbox{~for $T<0.3$}\\
      0.15\pi & \mbox{~for $0.3\leq T< 0.6$}\\
      \infty & \mbox{~for $T\geq 0.6$}\end{array}\right.
\end{equation}
If the annealing condition Eq.~(\ref{eq:annealing}) fail we repeat the
cooling with three times as many update sweeps; if it fails again we
increase the number of cooling sweeps a factor three again, and so on
until the condition is fulfilled. When ${\bf\Delta}_0$ is chosen,
before cooling, we let the system run until Eq.~(\ref{eq:annealing}) is
fulfilled with ${\bf\Delta}_0$ replaced by ${\bf\Delta}$.

 We repeat the above calculations for more
than 500 different disorder realizations (the disorder average is
denoted by $[ \cdots ]$ throughout this paper).

\begin{figure}
  \epsfxsize=8.2truecm
  \epsffile[60 310 540 650]{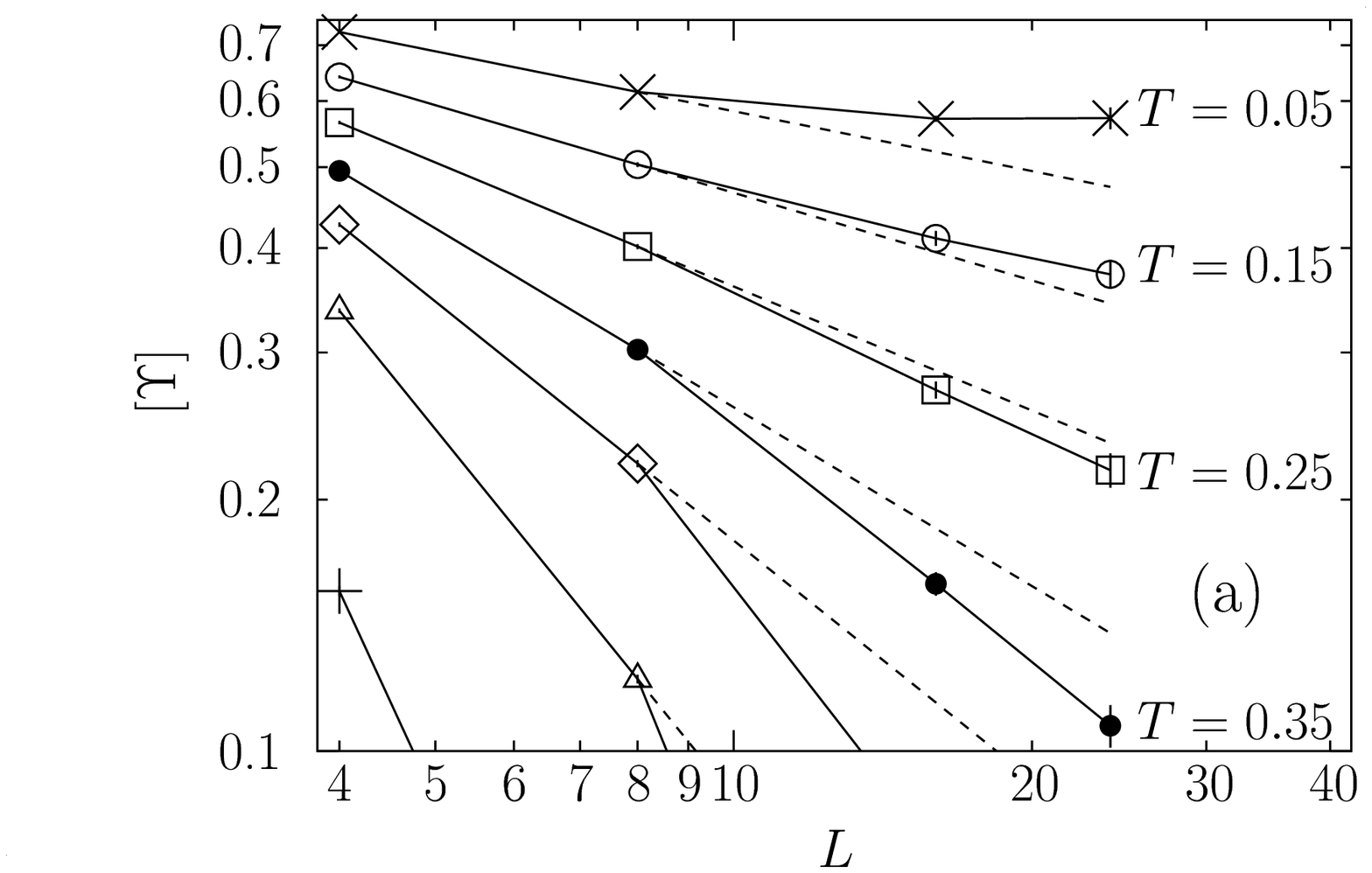}

  \epsfxsize=8.2truecm
  \epsffile[60 310 540 650]{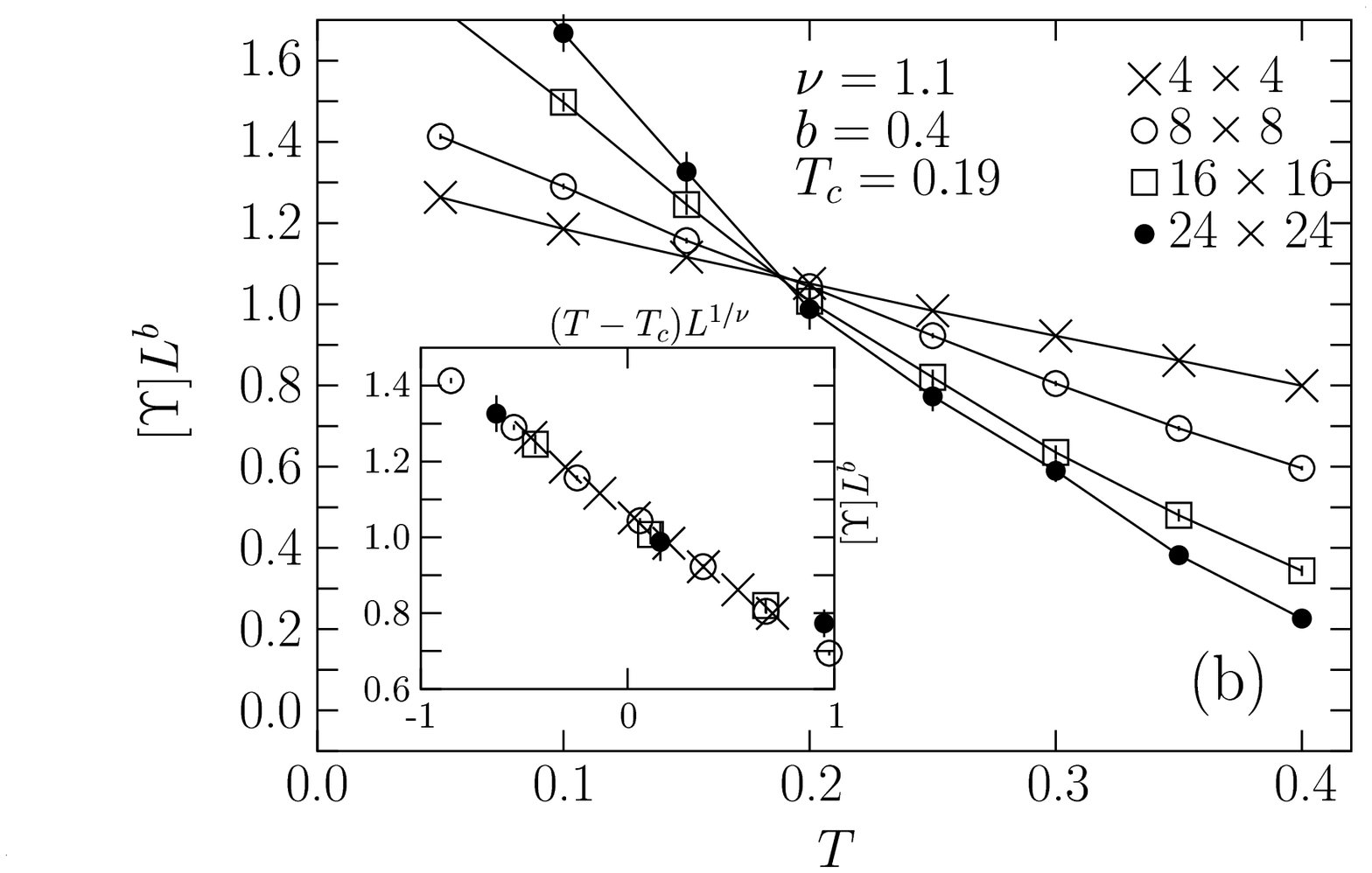}
\caption{(a) The size dependence of the helicity modulus $[
\Upsilon ]$ at different temperatures $T$ for the fully disordered
case ($r=1$)---the gauge glass model. At high $T$, $[\Upsilon]$ is
shown to vanish as $L$ is increased, whereas it saturates to a
finite value at low $T$ (qualitatively given by the dashed lines
are linear extrapolations from the smallest sizes and correspond
to power law behavior) (b) Finite-size scaling of $[\Upsilon]$ of
the data in (a). From the standard finite-size scaling form, the
critical temperature $T_c \approx 0.19$ and the critical exponent
$\nu \approx 1.1$ are determined.}\label{YL}
\end{figure}

For the case of $r=1$, we also use the periodic boundary condition PBC,
corresponding to $\mathbf{ \Delta} = 0$ in Eq.~(\ref{eq_H}), and compute
the root-mean-square current defined by~\cite{reger}
\begin{equation} \label{eq:irms}
\irms \equiv \left[ \left \langle \frac{1}{L}\sum_{\langle ij\rangle_x}
\sin \phi_{ij}  \right \rangle^2 \right]^{1/2} ,
\end{equation}
for comparisons with earlier works.

For $\irms$, for each disorder, we use 50000 update sweeps for
thermalizations and 500000 sweeps for measurements where the
actual measurement was performed every tenth sweep. We used this
computer time saving strategy since we found that performing the
measurement every sweep gave closely the same result.

\begin{figure}
  \epsfxsize=8.2truecm
  \epsffile[60 310 540 650]{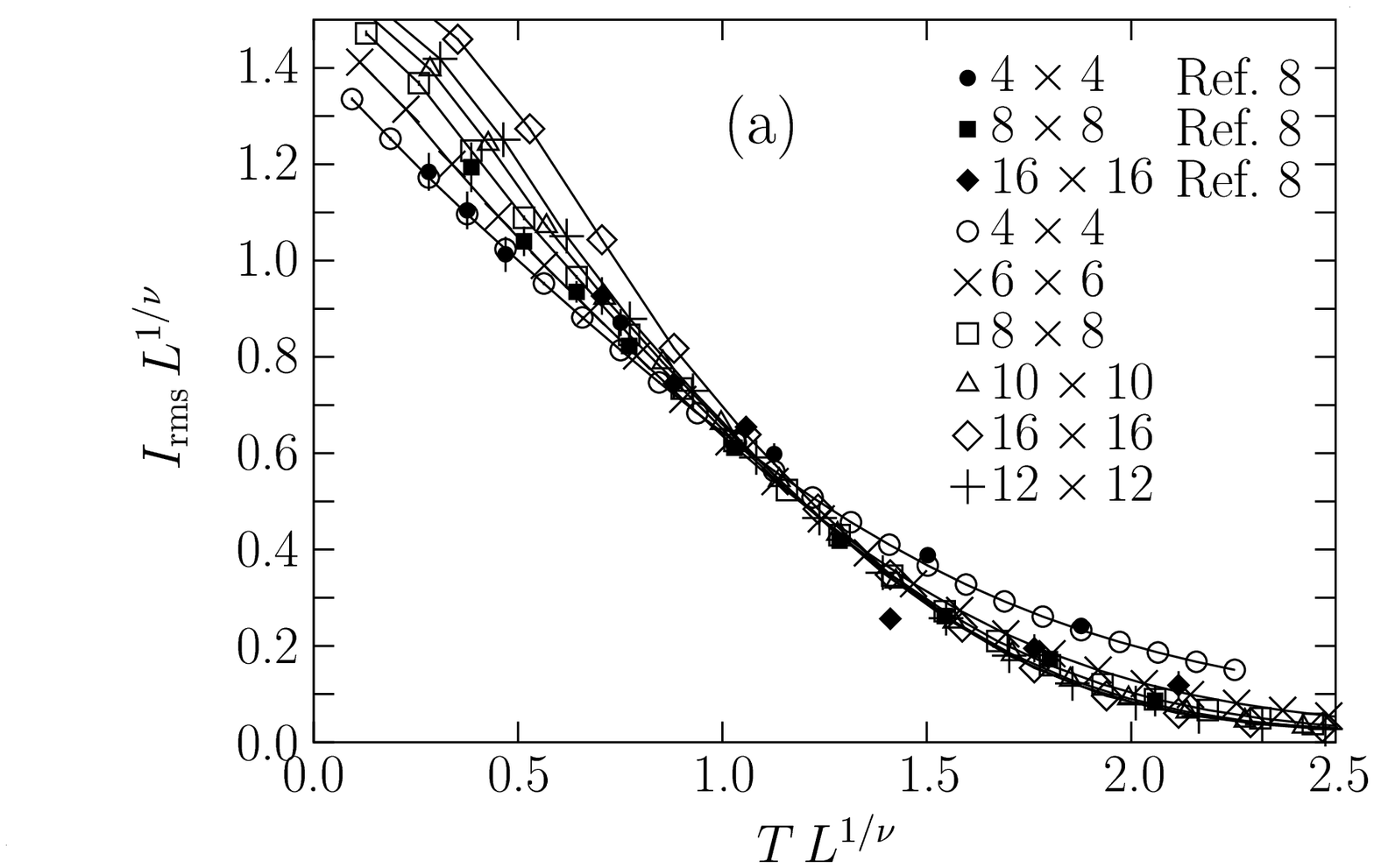}

  \epsfxsize=8.2truecm
  \epsffile[60 310 540 650]{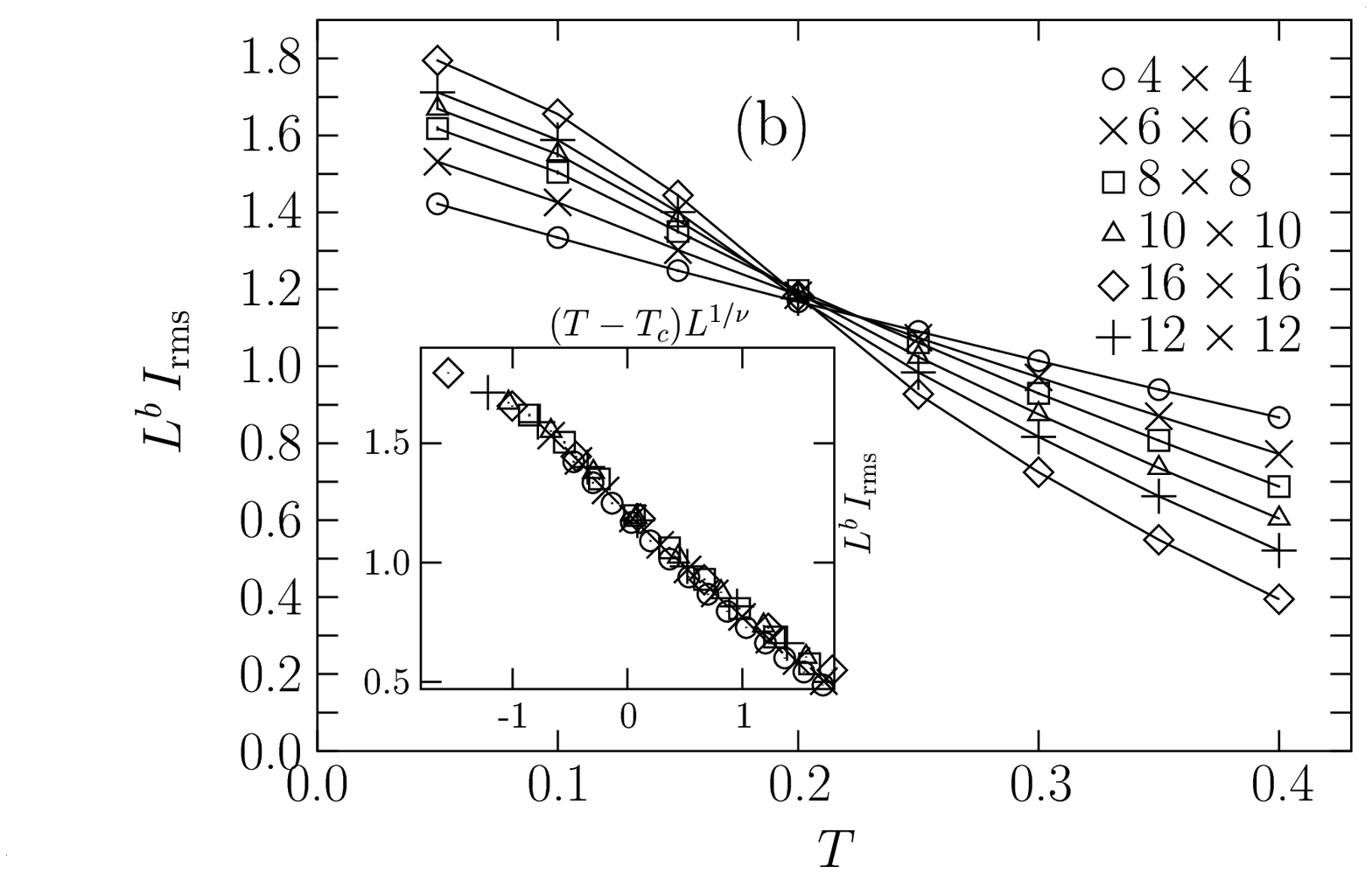}
\caption{The root-mean-square current $\irms$. (a) The finite-size-scaling
plot used in Ref.~\cite{reger} for $T_c = 0$
shows systematic deviations in the low-temperature region ($\nu = 2.2$ in
Ref.~\cite{reger} is used).
We include the data in Ref.~\cite{reger} for comparisons.
(b) Finite-size scaling form in Eq.~(\ref{eq:irmsfss}) yields $T_c \approx 0.2$
and $b \approx 0.5$. Inset: All data points in the main panel collapse
to a smooth curve with $\nu = 1.1$, the same value as found for the
helicity modulus in Fig.~\ref{YL}.
}
\label{fig:irms}
\end{figure}

\section{Results from Simulations}
We first investigate the standard 2D \xy gauge glass model,
corresponding to the fully disordered case ($r=1$), and show in
Fig.~\ref{YL}(a) the helicity modulus $[\Upsilon]$ as a function
of the system size $L$ for various temperatures. It is clearly
shown that at high $T$, $[\Upsilon]$ goes to zero as the system
size $L$ is increased. The crucial point is that, at low enough
temperatures ($ T \lesssim 0.2$), $[\Upsilon]$ changes its
curvature in terms of $L$, and appears to saturate to a finite
value as $L$ is increased. This behavior suggests a phase
transition with a scale-invariant power law dependence at the
critical temperature and a diverging characteristic length. Such a
behavior can often be described by the finite-size scaling form
\begin{equation} \label{eq:FSS}
[\Upsilon] = L^{-b} f\Bigl( L^{1/\nu} (T-T_c) \Bigr),
\end{equation}
where $[\Upsilon] \sim L^{-b}$ at the critical temperature $T_c$,
and the critical exponent $\nu$ is related to the divergence of
the coherence length. At $T_c$, the scaling function $f$ has the
same value irrespective of $L$, implying that $L^b [\Upsilon]$
versus $T$ should have a unique crossing point at $T_c$ for
various sizes. In Fig.~\ref{YL}(b), it is clearly shown that
$[\Upsilon]$ has a scale-invariant behavior $[\Upsilon]\propto
L^{-0.4}$ at a unique $T$, signaling a phase transition. The inset
of Fig.~\ref{YL}(b) furthermore confirms that this scaling
behavior is consistent with the standard form~(\ref{eq:FSS}) with
$\nu\approx 1.1$.

Simple dimensional analysis for the non-disordered case, $r=0$,
gives for $[\Upsilon]$ the exponent $b=0$ and from such a
dimensional analysis one would likewise conclude that $b=0$ also
for the disordered case. Thus if $[\Upsilon]$ scales with $b\neq
0$ for the disordered case this is equivalent to the appearance of
an anomalous dimension not accounted for by simple dimensional
analysis. Our suggestion, based on the simulation results, is
consequently that the disorder introduces such an anomalous
dimension.

Figure~\ref{fig:irms} shows the root-mean-square
current~(\ref{eq:irms}) for the PBC. We first note in
Fig.~\ref{fig:irms}(a) that the finite-size-scaling form for $T_c
= 0$ used in Ref.~\cite{reger}, $\irms = L^{-1/\nu} f( T L^{1/\nu}
)$ with $\nu = 2.2$, shows systematic deviations from the data
collapse to a single scaling curve at lower temperatures, in
contrast to what was concluded in Ref.~\cite{reger}, when more and
better converged data are included. Furthermore a $T_c=0$-collapse
cannot be achieved with any value of $\nu$. On the other hand if
we, in analogy with the finding for $[\Upsilon]$ above, use the
scaling form
\begin{equation} \label{eq:irmsfss}
\irms =  L^{-b} f\Bigl( L^{1/\nu} (T-T_c) \Bigr),
\end{equation}
which allows for the same anomalous dimension, one obtains the
scaling plot in Fig.~\ref{fig:irms}(b)  with $b \approx 0.5$ and
$T_c \approx 0.2$.\cite{note} From Figs.~\ref{YL} and
\ref{fig:irms} we conclude that the 2D $XY$ gauge glass ($r=1$)
exhibits a non-KT type superconducting to normal transition at
$T_c \approx 0.2$ characterized by the existence of the anomalous
dimension $b \approx 0.5$ and $\nu \approx 1.1$.

Next we investigate the phase transition with decreasing $r$.
For $r=0.9,0.8,\cdots ,0.5$,
we obtain the finite-size scaling plots of the same quality as in
Fig.~\ref{YL}(b), and determine the phase boundary
(the dashed line in Fig.~\ref{PhD}).
As $r$ is changed
from $0.5$ to $0.4$ the exponents $b$ and $\nu$ exhibit quite
rapid changes, $b$ from 0.27 to 0.06 and $\nu$ from 1.1 to 2.0.
This, in our interpretation, reflects that near $r = 0.4$
the nature of the superconducting-normal transition changes from a
non-KT type to a KT type.

\begin{figure}
  \epsfxsize=8.2truecm
  \epsffile[60 310 540 650]{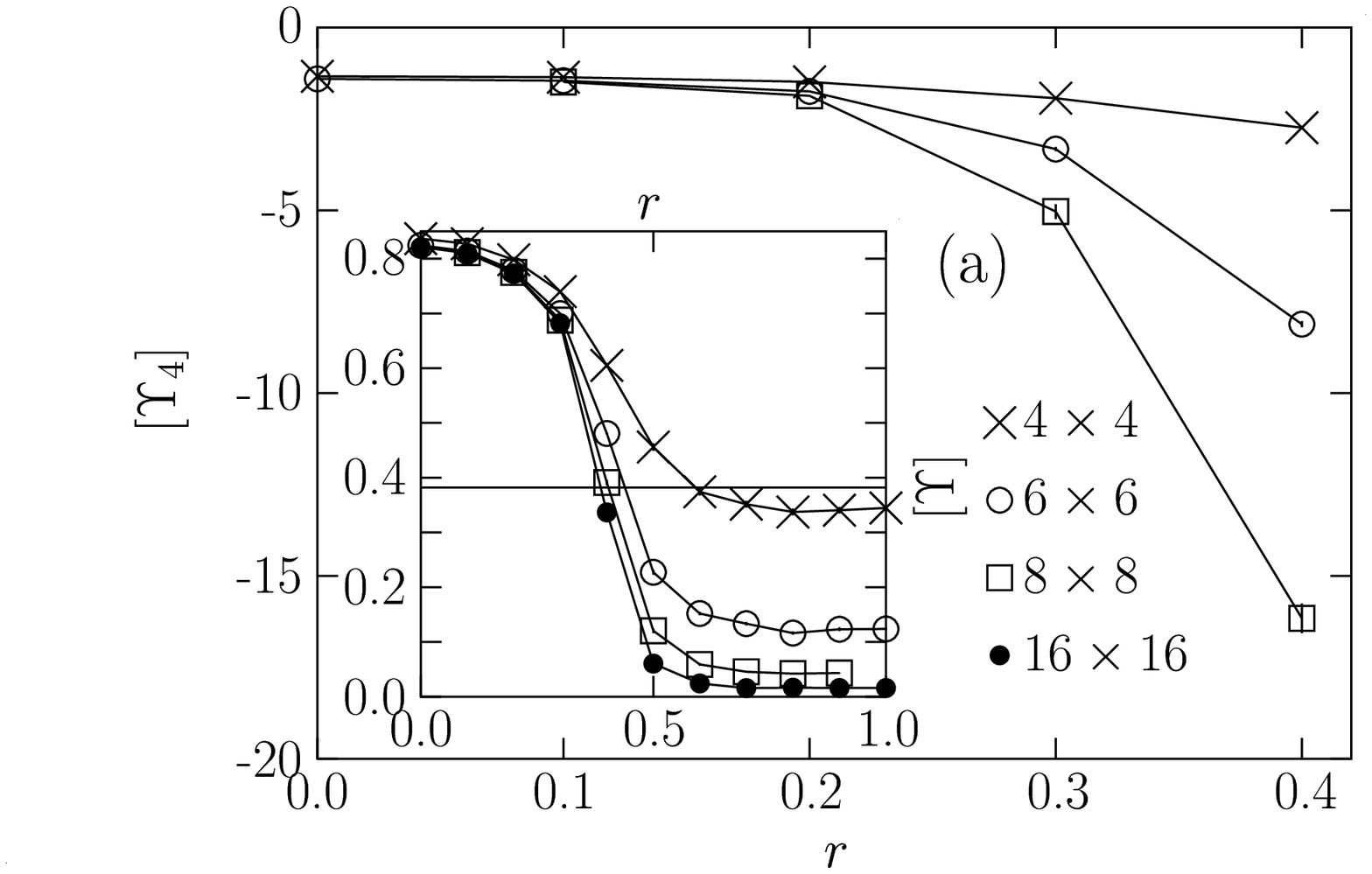}

  \epsfxsize=8.2truecm
  \epsffile[60 310 540 650]{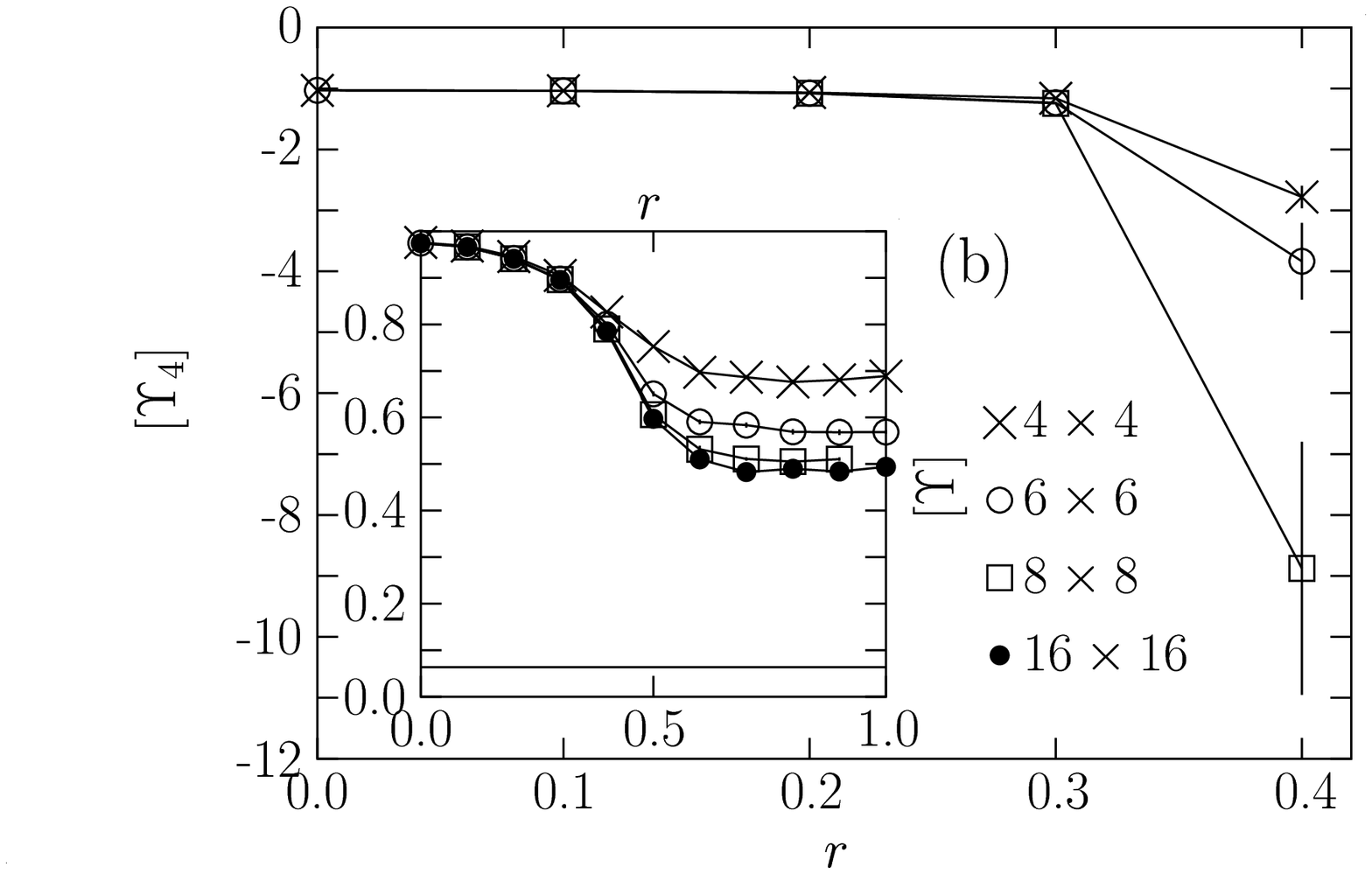}
\caption{The fourth order modulus and helicity modulus
  (inset) as a functions of $r$  for temperatures (a) 0.6 and (b) 0.1.
  The solid line in the inset represents the universal-jump condition
  for a KT transition $[\Upsilon]=2T/\pi$.}
\label{YY}
\end{figure}

For $r=0$ the phase transition is of the KT nature and it has been suggested
 that this character should persist along the phase boundary  up to some
$r_c<1$~\cite{maucourt,more}. The KT transition is characterized
by that $[\Upsilon]$ jumps from a finite value $[\Upsilon]=2T/\pi$
to zero as the phase line is crossed from the small ($T$,$r$) region.
For $r=0$ we find that
the KT transition is also characterized by the increase of the
4th order modulus $|[\Upsilon_4]|\propto L^c$
at $T_c$ with a positive exponent $c$, and $[\Upsilon_4]$ stays at
a constant value below $T_c$ as $L$ is increased~\cite{gang}.
Figure~\ref{YY}~(a) shows $[\Upsilon_4]$ for $T=0.6$ as a function of $r$.
The KT condition in the inset gives $r\approx 0.4$.
This means that if the transition is of KT type then it
would occur around $r\approx 0.4$. The onset of size dependence for
$[\Upsilon_4]$ in Fig.~\ref{YY}~(a) is consistent with a transition at around
$r\lesssim 0.4$. Thus the transition is compatible with a
 KT-character, although a different character cannot be ruled out. The situation
 for $T=0.1$ in Fig.~\ref{YY}~(b) is quite different: the inset shows
 that the KT jump condition is not fulfilled and a KT transition can be ruled out.
 Yet there is a marked structure in $[\Upsilon]$ around $r\approx
 0.4$. This structure corresponds to the onset of strong size
 dependence in $[\Upsilon_4]$. We interpret this onset as the
 reflection of a true divergence in $[\Upsilon_4]$ consistent with a phase
 transition. Thus we suggest that the whole boundary to the low-$T$ KT
 phase (solid line in Fig.~\ref{PhD}) is reflected in a divergence of
 $[\Upsilon_4]$ and that this line ends at ($T=0$,$r_c\approx 0.4$).
 A phase line which ends at ($T=0$, $r_c\approx 0.4$) has been found in
 many earlier investigations (see e.g. Ref.~\cite{maucourt}
 and references therein)\cite{new}. The difference with earlier work is that in
 our case such a phase line
 is for lower temperatures between the two
 different superconducting phases SI and SII (as demonstrated by our
 direct calculation of $[\Upsilon]$), whereas earlier work have
 concluded that such a phases line does indeed exist but separates a superconducting phase
 from a normal phase all the way down to $T=0$. Our suggested phase
 diagram is thus
 consistent with earlier work as to the existence of a phase line
 ending at ($T=0$, $r_c\approx 0.4$).
 For temperatures larger than the merging with the second phase line
 (above dashed line in Fig.~\ref{PhD}) the transition is consistent
 with a KT transition although a different character cannot be ruled
 out. However, below the merging with this second line the transition is not a KT transition.
 It may be that there still is a jump in  $[\Upsilon]$ at this
 transition, but this is then between two non-vanishing values.

Based on the numerical evidences we suggest the structure of the
phase diagram is sketched in Fig.~\ref{PhD}. One striking feature
is the finite-$T$ transition line between normal and
superconducting phases, starting from the boundary of the low-$T$
KT-phase (solid line) and changing into the (dashed) line which
ends at $T_c \approx 0.2$ at $r=1$. The latter line (dashed line)
is characterized by the appearance of an anomalous dimension $b$.
Since the KT transition does not have such an anomalous dimension
it follows that, if the transition along the boundary to the
low-$T$ KT phase boundary has KT character, then
 $b$ should approach $b=0$ when
the two transition line merge. Thus
the rapid drop from $b \approx 0.27$ to $b \approx 0.06$
as $r$ is changed from $r=0.5$ to 0.4 is consistent with a change over to a
KT transition.
Another interesting feature is that the phase line (solid line
in Fig.~\ref{PhD}), associated with the divergence of $[\Upsilon_4]$,
continues even inside the superconducting region and ends at $r \approx 0.4$
for $T=0$. Although the transition separating the KT phase and
the normal phase (SI and N, respectively, in Fig.~\ref{PhD}), may well be of the true
KT type, manifested by a universal
jump in the helicity modulus,
the phase line separating SI and SII in Fig.~\ref{PhD}
is not a KT transition and the helicity modulus has
a non-zero value on both sides.

It is interesting to note that earlier works have found evidences
for only two phases separated by a \textit{ single} phase line;
either, in the more prevailing view, a phase line ending at a point $T=0$ for a finite $r_c<1$,
or, in the less prevailing view, a phase line ending at a point $T>0$ for $r=1$.
>From our numerical simulations we instead suggest three distinct phases
separated by $two$ phase lines,
which combines the two earlier proposed scenarios and provides
a unified picture:
On the one hand, in Ref.~\cite{simkin}, the
end point of the phase boundary to the low-$T$ KT-phase has been obtained
to be $r_c\approx 0.37$ and $T=0$,  which is
consistent with our cruder estimate $r\approx 0.4$.
On the other hand, the phase transition point
at $T \approx 0.22$ with $\nu  = 1.1$, found in
Refs.~\cite{choi} and \cite{bjk} for $r=1$,
is in very good agreement with the end point of our finite-$T$ phase line
($T\approx 0.2$, $\nu\approx 1.1 $).
However, the difference with the previous work~\cite{simkin}
is that according to our interpretation the phase line below $T \approx 0.2$
ending at $T=0$ and $r_c\approx 0.4$
separates two distinct superconducting phases, while the whole phase line in
Ref.~\cite{simkin} is for the superconducting to normal transition. A
very hand-waving picture of the scenario of our phase diagram in terms of
vortex motion is sketched in Fig.~\ref{PhD}: In SI the vortex motion is
suppressed by vortex pair binding, in N pair unbinding has occurred and
free vortices exists which are not entirely pinned by the disorder,
whereas in SII vortex pair unbinding has occurred but the vortices are
pinned by the disorder.

\section{Summary}
One main conclusion from this phase diagram is that the \xy gauge
glass model ($r=1$) has a finite-$T$ transition. How is this
possible in view of earlier conflicting evidences? In
Ref.~\cite{reger} it was concluded, on the basis of an analysis of
data for the root-mean-square current for standard periodic
boundary conditions~\cite{foot}, that no finite-$T$ transition
exists in 2D. We have found that, taking into account the
possibility of an anomalous scaling dimension, $\irms$ displays a
transition at a finite-temperature (see
Fig.~\ref{fig:irms})~\cite{gang}. Another puzzling evidence to the
contrary is the $T=0$ calculations of the size scaling of the
defect energy~\cite{3d,simkin,other:T0:studies,FTY}. However, as
discussed in Refs.~\cite{FTY} and \cite{holme}, the local
vorticity conservation must be properly taken into account when
calculating the energy barriers. Thus the energy barrier for
vortex dissipation increases with system size when taking the
local vorticity conservation into account\cite{holme}. This
growing of the energy barrier for vortex dissipation with system
size supports the possibility of a finite-$T$
transition~\cite{holme}.

The appearance of a new superconducting phase for the \xy random gauge
model is intriguing.
In particular, since it is neither a low $T$ KT phase nor a phase with
a finite glass order parameter.
The true nature of this phase and the existence of similar phases
in other related models are open questions.

Support from the Swedish Natural Research Council through Contract
No.\ F 5102-659/2001 is gratefully acknowledged.
B.J.K. was  supported by the Korea Science and Engineering
Foundation through Grant No. R14-2002-062-01000-0.


\begin{thebibliography}{99}
\bibitem{huse} D.~A.~Huse and H.~S.~Seung, Phys.\ Rev.\ B \textbf{42}, 1059
  (1990).
\bibitem{fisher} M.~P.~A.\ Fisher, Phys.\ Rev.\ Lett.\ \textbf{62}, 1415 (1989);
  D.~S.~Fisher, M.~P.~A.~Fisher, and D.~A.~Huse, Phys.\ Rev
  B \textbf{43}, 130 (1991).
\bibitem{3d}
  M.~Cieplak, J.~R.~Banavar, and A.~Khurana, J.\ Phys.\ A \textbf{3}, L145
(1991);
  M.~Cieplak \textit{ et al.}, Phys.\ Rev.\ B \textbf{45}, 786 (1992).
\bibitem{noT0}
  J.~D.~Reger \textit{ et al.}, Phys.\ Rev.\ B \textbf{44}, 7147 (1991);
  C.~Wengel and A.~P.~Young, Phys.\ Rev.\ B \textbf{56}, 5918 (1997).
\bibitem{simkin}
  J.~M.~Kosterlitz and M.~V.~Simkin, Phys.\ Rev.\ Lett.\ \textbf{79}, 1098
(1997).
\bibitem{other:T0:studies}
  M.~J.~P.~Gingras, Phys.\ Rev.\ B \textbf{45}, 7547 (1992);
  H.~S. Bokil and A.~P. Young, Phys.\ Rev.\ Lett.\ \textbf{74}, 3021
(1998);
  J.~M.~Kosterlitz and M.\ Akino, Phys.\ Rev.\ Lett.\ \textbf{81}, 4672
(1998).
\bibitem{FTY}
  M.~P.~A.\ Fisher, T.~A.\ Tokuyasu, and A.~P.\ Young,
  Phys.\ Rev.\ Lett.\ \textbf{66}, 2931 (1991).
\bibitem{reger} J.~D.\ Reger and A.~P.\ Young, J.\ Phys.\ A \textbf{26}, L1067
  (1993).
\bibitem{hyman} R.~A.\ Hyman \textit{ et al.}, Phys.\ Rev.\ B \textbf{51}, 15304 (1995).
\bibitem{nishimori} H.~Nishimori, Physica A \textbf{205}, 1 (1994).
\bibitem{choi} M.~Y.\ Choi and S.~Y.\ Park, Phys.\ Rev.\ B \textbf{60},
  4070 (1999).
\bibitem{li} Y.~H.\ Li, Phys.\ Rev.\ Lett.\ \textbf{69}, 1819 (1992).
\bibitem{bjk} B.~J.\ Kim, Phys.\ Rev.\ B \textbf{62}, 644 (2000).
\bibitem{bjk1} B.~J.\ Kim \textit{ et al.}, Phys.\ Rev.\ B \textbf{56},
6007 (1997).
\bibitem{jeon} G.~S.\ Jeon, S.~Kim, and M.~Y.\ Choi, Phys.\ Rev.\ B
  \textbf{51}, 16211 (1995).
\bibitem{maucourt} J.\ Maucourt and D.~R.\ Grempel, Phys.\ Rev.\ B
  \textbf{56}, 2572 (1997).
\bibitem{more} M.~Rubinstein, B.~Shraiman, and D.~R.\ Nelson, Phys.\ Rev.\ B
  \textbf{27} 1800 (1983); T.\ Natterman \textit{et al.}, J.~Phys.\ (France) I
  \textbf{5}, 565 (1995); M.-C.\ Cha and
  H.~A.\ Fertig, Phys.\ Rev.\ Lett.\ \textbf{74}, 4867 (1995); L.-H.\ Tang,
  Phys.\ Rev.\ B \textbf{54}, 3350 (1996).
\bibitem{note} Preliminary results for parallel tempering Monte
Carlo give the same qualitative result.\cite{gang}
\bibitem{olsson} P.~Olsson, Phys.\ Rev.\ B \textbf{46}, 14598 (1992);
  \textbf{52}, 4526 (1995).
\bibitem{gang} For more details, B.~J.\ Kim, P.~Minnhagen, and P.~Holme
  (unpublished).
\bibitem{new} This boundary of SI is completely in accord with the change
  from power law to exponential decay of the spin correlation function
  found in Ref.\onlinecite{maucourt}\cite{gang}. In SII the higher order
  correlation function discussed in Ref.\onlinecite{jeon} has a
  power-law decay\cite{gang}.
\bibitem{foot} With the FTBC, the r.m.s.\ current is identically zero.
\bibitem{holme} P.~Holme and P.~Olsson, Europhys.\ Lett.\ \textbf{60}, 439
(2002).

\end{thebibliography}
\end{document}